\documentclass[aps,prl,twocolumn,superscriptaddress,longbibliography]{revtex4-1}

\usepackage{amsthm}
\usepackage[utf8]{inputenc}
\usepackage{amsmath,amssymb,amsmath,amsthm,bm,graphicx,xcolor}
\usepackage{natbib}
\usepackage[breaklinks]{hyperref}
\usepackage{color}
\usepackage{tabularx}
\usepackage{epsfig}
\usepackage{amssymb}
\usepackage{graphicx}
\usepackage{ytableau}
\usepackage{wasysym}
\usepackage{nicefrac}
\usepackage{dcolumn}
\usepackage{epstopdf}
\usepackage{subfigure}
\usepackage{psfrag}
\usepackage{bbold}
\usepackage[breaklinks]{hyperref}
\hypersetup{colorlinks=true}

\usepackage{color}
\usepackage{tabularx}
\usepackage{epsfig}
\usepackage{graphicx}
\usepackage{wasysym}
\usepackage{dcolumn}
\usepackage{epstopdf}
\usepackage{subfigure}
\usepackage{latexsym}
\usepackage{psfrag}
\usepackage{cleveref}

\newcommand{\rr}{\bm{r}}

\usepackage{physics}
\begin{document}
\author{Johannes Gedeon}
\author{Jonathan Schmidt}
\affiliation{Institut f\"ur Physik, Martin-Luther-Universit\"at
Halle-Wittenberg, 06120 Halle (Saale), Germany}
\author{Matthew J. P. Hodgson}
\affiliation{Department of Physics, Durham University, South Road, Durham, DH1 3LE, United Kingdom}
\author{Jack Wetherell}
\affiliation{LSI, \'Ecole Polytechnique, CNRS, Institut Polytechnique de Paris, F-97728 Palaiseau}
\author{Carlos L. Benavides-Riveros}
\affiliation{Max Planck Institute for the Physics of Complex Systems, N\"othnitzer Str.~38, 01187, Dresden, Germany}
\affiliation{NR-ISM, Division of Ultrafast Processes in Materials (FLASHit), Area della Ricerca di Roma 1, Via Salaria Km 29.3, I-00016 Monterotondo Scalo, Italy}
\author{Miguel A. L. Marques}
\email{miguel.marques@physik.uni-halle.de}
\affiliation{Institut f\"ur Physik, Martin-Luther-Universit\"at
Halle-Wittenberg, 06120 Halle (Saale), Germany}

\title{Machine learning the derivative discontinuity of density-functional theory}

\date{\today}

\begin{abstract}
Machine learning is a powerful tool to design accurate, highly non-local, ex\-chan\-ge-correlation functionals for density functional theory. So far, most of those machine learned functionals are trained for systems with an integer number of particles. As such, they are unable to  re\-pro\-du\-ce  some  crucial  and  fundamental  aspects, such as the explicit dependency of the functionals on the particle number or the infamous derivative discontinuity at integer particle numbers. Here we propose a solution to these problems by training a neural network as the universal functional of density-functional theory that (i)~depends explicitly on the number of particles with a piece-wise linearity between the integer numbers and (ii)~reproduces the derivative discontinuity of the exchange-correlation energy. This is achieved by using an ensemble formalism, a training set containing fractional densities, and an explicitly discontinuous formulation.
\end{abstract}
\maketitle
\section{Introduction}
In their now famous paper, Hohenberg and Kohn pro\-ved that the electron density $\rho(\rr)$ suffices to compute all observables of a system of interacting electrons~\cite{HohenbergKohn}. Due to a remarkable balance of computational cost and numerical precision, first principles modeling of electronic systems based on this density functional theory (DFT) is nowadays a daily practice, with great impact in  material science, quantum chemistry or condensed matter~\cite{Jones}.  The success of DFT is to a large extent based on the Kohn-Sham formulation, that utilizes a system of non-interacting electrons that has the same density as the interacting one~\cite{KohnSham}. The main ingredient of this formulation is $E_{\rm xc}[\rho]$, the universal exchange-correlation (xc) functional,  whose functional derivative provides an effective external potential for the non-interacting particles. Yet, while the Hohenberg-Kohn theorem proves the uniqueness of such a functional, it does not give any indication regarding its specific form. To circumvent this issue, a very large number of approximate functionals were developed in the last decades~\cite{libxc, Perdew2001}, often combining empirical knowledge, exact mathematical conditions, and a great deal of ingenuity. 

Inspired by the success of machine learning (ML) in va\-rious technological applications, including image and speech recognition~\cite{Deng}, the last couple of years have seen the development of several neural-network-based approximations to $E_{\rm xc}[\rho]$.
Indeed, it is now firmly established that machine-learning offers a new generation of  accurate, highly non-local, xc functionals~\cite{Kalita2021}. While those functionals are designed to perform tasks of different degree of complexity, all share the aim of learning one of the maps of DFT, namely, the Hohenberg-Kohn map between the external potential $v(\rr)$ and the density $\rho(\rr)$ \cite{Brockherde2017,KSRegulizer,Margraf2021,PhysRevLett.125.076402,PhysRevA.100.022512}, or the Kohn-Sham map between the density $\rho(\rr)$ and the xc functional $E_{\rm xc}[\rho(\rr)]$ and its functional derivative $v_{\rm xc}[\rho(\rr)]=\delta E_{\rm xc}[\rho(\rr)]/\delta \rho(\rr)$~\cite{Schmidt2019,Nagai2020,Neupert2020}. 

The functionals delivered by machine-learning DFT (ML-DFT) are in general non-local, in the sense that they use multiple density points as input, and can be efficiently trained with data from reference methods. Yet, since ML-DFT functionals are mostly trained in Hilbert spaces with an \textit{integer} number of particles, they are still unable to reproduce some critical and fundamental aspects of DFT. For instance, it is known that any satisfactory definition of the energy functional must depend explicitly on the particle number \cite{Lieb,doi:10.1142/9789812775702_0022,PhysRevLett.110.126403}. Furthermore, the derivative of the xc functional in terms of the number of particles exhibits a discontinuity that plays a crucial role in the description of electronic bandgaps~\cite{Sham1983,Perdew1982,PhysRevB.74.161103,PhysRevLett.107.183002}, charge-transfer excitations~\cite{Hodgson2017,PhysRevLett.111.126402}, molecular dissociation~\cite{C4CP01170H,Wasserman2014,Baerends2019,Perdew1985}, or even Mott insulators~\cite{10.1021/cr200107z}, to name but a few examples. 

Systems with non-integer (fractional) number of electrons ($N + \epsilon$) are defined as statistical mixtures of systems with integer number of particles~\cite{Perdew1982,Yang2000}. As such, the density $ \rho_{N+\epsilon}(\rr)$ and total energy $E(N+\epsilon)$ are piecewise linear functions of $\epsilon$, namely:
\begin{subequations}
\begin{align}
  \rho_{N+\epsilon}(\rr) &=(1-\epsilon)\rho_N(\rr)+\epsilon \rho_{N+1}(\rr), \\
  E(N+\epsilon) &=(1-\epsilon)E(N)+\epsilon E(N+ 1)\,
\end{align}
\end{subequations}
with $0\leq \epsilon\leq 1$. At integer $N$ (i.e., when $\epsilon = 0$), the derivatives of the density and the energy exhibit a discontinuity, and the xc potential $v_{\rm xc}(\rr)$ jumps by a finite value~\cite{PhysRevLett.51.1884}.
The difference in the slope on the left/right side of the total energy at integer values is equal to the fundamental gap~\cite{Perdew1982}:
\begin{equation}\label{jumpXC}
 I-A = \left.\frac{\partial E}{\partial N}\right|_{+}-\left.\frac{\partial E}{\partial N}\right|_{-}\,, 
\end{equation}
where $I$ is the ionization energy and $A$ the electron affinity. Yet, in practice, standard approximations to the xc functionals that depend explicitly on the electronic densi\-ty, such as the local-density (LDA) and generalized-gradient (GGA) approximations, are continuously differentiable functions of $N$ and lack therefore a derivative discontinuity. Meta-GGAs can exhibit a discontinuity due to their dependence on the kinetic-energy density, but it is usually too small or even negative~\cite{Eich}. Due to their dependence on the Kohn-Sham orbitals, orbital functionals are discontinuous~\cite{Kuemel2008}, but this comes at the price of a much higher computational effort.

In addition to the discontinuity, a universally useful approximation for the xc functional must be  ``$N$-electron self-interaction-free'' for \textit{all} posi\-ti\-ve integer $N$ \cite{Ruzsinszky2006}, meaning that the total energy of  a  system  with $N +\epsilon$ electrons in the range $(N,\,N+1)$ should exhibit a linear  variation with respect to $\epsilon$. For attractive interactions the energy is a convex function with straight lines joining subsets of ground-state energies \cite{perfetoo20212}. Yet, approximate functionals deviate from such a correct behavior. It has been shown that semi-local density functionals are in general convex with perhaps small concave pieces \cite{Li2017}. Even the Hartree Fock theory leads to piecewise concave curves between integers \cite{Li2017}. We note that the relatively well-defined curvature of the curves is ultimately the reason for the success of the Slater half-occupation scheme \cite{PhysRevB.5.844} or the LDA-$1/2$ method \cite{PhysRevB.78.125116}. In fact, these schemes use the derivative at the midpoint (i.e., at $N-0.5$), that can be shown to be equal to the slope of the straight line between $E(N-1)$ and $E(N)$ if the curvature is constant, irrespective of its sign~\cite{Baerends2018}.
The centrality of these pressing issues in DFT can be further highlighted by the fact that a rigorous description of the delocalization error can be related with the energy curve of the xc functionals lying below the straight energy lines \cite{Li2017,Martin2018}. 

In this work, we propose a way to train a neural network as the \textit{ensemble} universal functional of a system of fractional electron numbers that describes correctly the derivative discontinuity and the piecewise linear behavior. The ML functionals we present contain explicitly the physics of the derivative discontinuity of DFT, are highly non-local, and are trained for systems with fractional densities. For this reason, our functionals can potentially address the well known delocalization and static correlation errors of DFT \cite{Cohen792,PhysRevLett.100.146401,C7CP01137G} simultaneously.

\section{Results and discussion}

Inspired by the neural network topology proposed in Ref.~\cite{Schmidt2019}, our neural network takes an electronic density as an input and returns the corresponding xc energy, whose functional derivative can in turn be used to solve the Kohn-Sham equations. The network is a sliding window convolution (SWC) network. For a 1D system of discrete spatial points $\{r_1,...,r_W\}$, a \textit{window} with a certain \textit{kernel size} $\kappa$ scans each data point $\rho_\sigma(r_j)$ and its $\kappa-1$ nearest neighbors $\eta_{\sigma}(r_j,\kappa) = \{\rho_\sigma(r_{j-(\kappa-1)/2}),...,\rho_\sigma(r_{j+(\kappa-1)/2})\}$ with $\sigma = \uparrow, \downarrow$ to calculate a local energy $ \epsilon^{\theta}_{ \mathrm{loc}}\big[\eta_{\uparrow}(r_j,\kappa),  \eta_{\downarrow}(r_j,\kappa)\big]$. The total xc energy is calculated by summing over the local energies:
\begin{equation}\label{ExcOuput}
E_{\mathrm{xc}}\left[\rho_{\uparrow}, \rho_{\downarrow}\right]=
\sum_j \rho(r_j) \, \epsilon^{\theta}_{ \mathrm{loc}}\big[\eta_{\uparrow}(r_j,\kappa),  \eta_{\downarrow}(r_j,\kappa)\big] \,.
\end{equation}
Here, $\theta$ denotes the trainable parameters. The input channels can be the total electronic density $\rho = \rho_\uparrow + \rho_\downarrow$ or the spin densities $\rho_\uparrow$ and  $\rho_\downarrow$.
The corresponding xc potentials can be computed using automatic differentiation, as shown in Ref.~\cite{Schmidt2019}.  The parameters of the neural network are updated according to the loss function:
\begin{align}
\mathcal{L}(\theta;\alpha,\beta) &= \alpha \,\mathrm{MSE}(v^{\uparrow}_{\mathrm{xc}}[\rho_{\uparrow}, \rho_{\downarrow}],v^{\downarrow}_{\mathrm{xc}}[\rho_{\uparrow}, \rho_{\downarrow}] ) \nonumber \\ & \qquad +\beta \label{LossFunction1} \,\mathrm{MSE}(E_{\mathrm{xc}}[\rho_{\uparrow}, \rho_{\downarrow}])\,,
\end{align}
where $\alpha$ and $\beta$ are \textit{fixed} weights that can be adjusted to expedite convergence, and MSE is the mean squared error. A more detailed description of our networks can be found in the Methods section.

We improve the per\-for\-mance of this architecture by (i) training our neural network with non-integer densities, (ii) introducing the jump of the xc potential at integer numbers into the loss function, and  (iii) adding an explicit discontinuity at integer electron numbers. In the following we discuss the details of these new approaches. 

\subsection{Fractional particle numbers}

\begin{figure}[t!]
\includegraphics[width=0.85\linewidth]{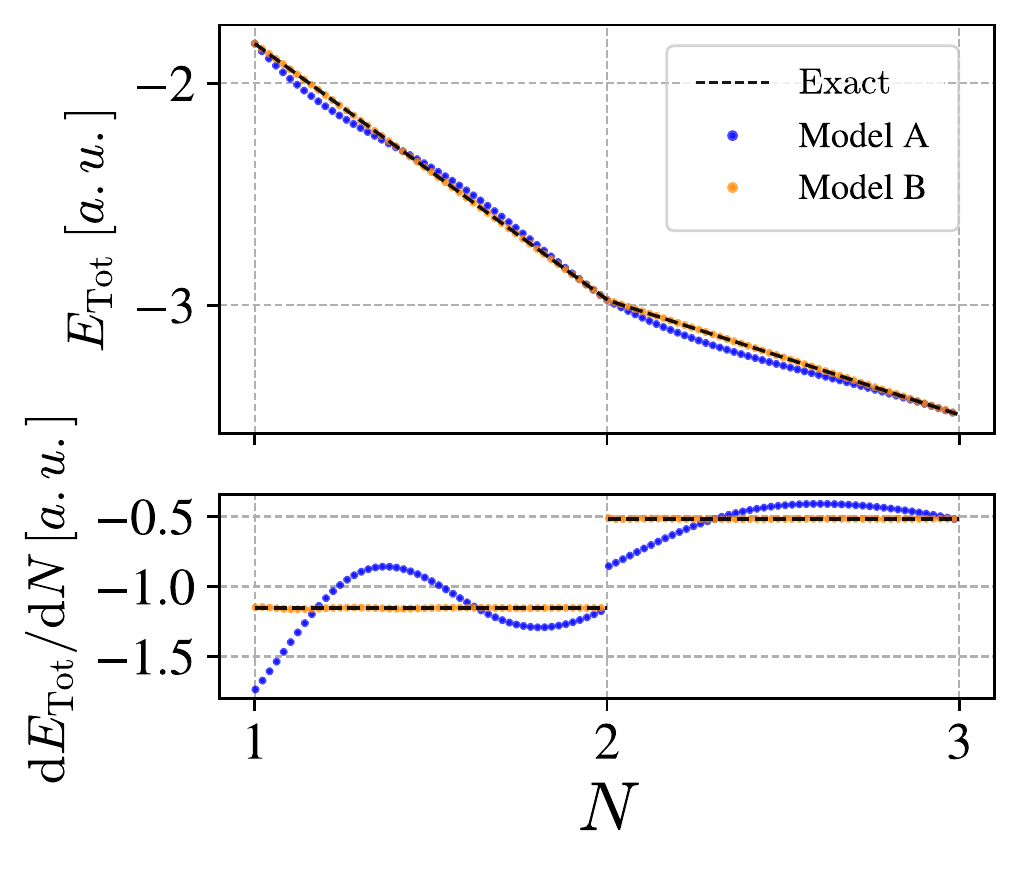}
\caption{\label{BestModel}
Comparison of the total energy for a certain external potential (top) as well as its derivative with respect to the particle number (bottom) for two different models: Model A has been trained with integer densities only. Model B has been trained with fractional densities, the exact $\Delta_{\mathrm{xc}}$-shift, and employs the AF (see text).
}
\label{Pic:EnergyCurve}
\end{figure}

In general, neural networks do not extrapolate well outside the distribution of the samples used for their training . Consequently, we can not expect that machines trained solely for integer densities (as usually done) will exhibit the correct linear behaviour of the energy. To illustrate this behavior we plot, in Fig.~\ref{Pic:EnergyCurve}, the total energy calculated with a neural network trained solely with integer densities (model A in the figure) as a function of the number of particles for a 1-dimensional system. Besides the fact that model A is far from linear, we can also note that the sign of the curvature is not constant, with both concave and convex parts. This is somehow to be expected, as the network, in contrast to the usual xc functionals, only incorporates physical knowledge through the training examples with integer densities. As such, we can easily see that approaches such as LDA-$1/2$ are bound to fail in this case.

As a first strategy to solve this problem, we decided to include samples calculated within ensemble-DFT at fractional densities in our training. To obtain this data we created a set of total energies and electronic densities for a series of 1-di\-men\-sional exact calculations. We then constructed ensemble densities and inverted the ensemble Kohn-Sham equations, in order to compute the exact xc energy and potential for these systems. We used an inversion algorithm based on Ref.~\cite{Inversion} that we ex\-ten\-ded for both spin-DFT \cite{Barth_1972} and to ensemble systems. As a result, we created exact training and testing data with particle numbers between 1 and 3 electrons, that we used to train our models.

\begin{figure}[t!]
\includegraphics[width=0.9\linewidth]{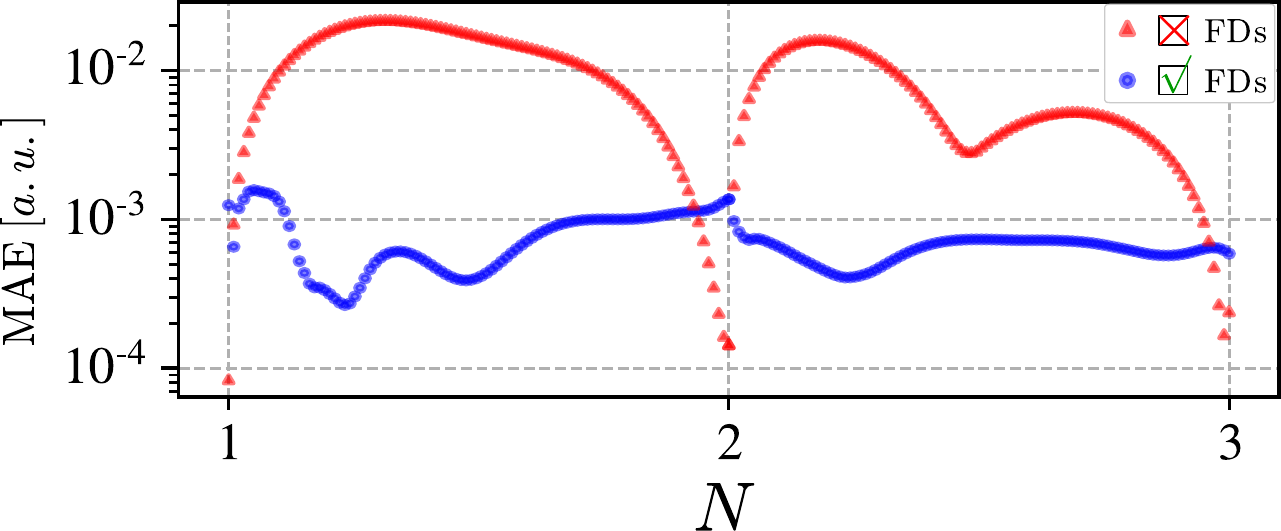}
\caption{\label{Pic:CompareModels2}
Comparison of the mean absolute error (MAE) of the total energy between two different models (averaged over 100 external potentials). The model ``FDs'' (blue) includes fractional densities $\epsilon= (0.05, 0.2, 0.5, 0.8, 0.95)$ in the training. The second model (red) includes only integer densities in the training. Both models were trained within the spin-DFT framework.}
\end{figure}

In Fig.~\ref{Pic:CompareModels2} we compare the mean absolute error (MAE) of the total energy for a functional trained with fractional densities (blue) and a functional trained only for systems with integer densities (red) (averaged over a test set of 100 external potentials). As we can see, the model trained at integer densities yields an excellent prediction for the total energy at those integers, but exhibits a considerably larger error at fractional numbers. Remarkably, by simply adding fractional densities to the training set, the error decreases by more than one order of magnitude, and the MAE over the entire $[1,3]$-range remains below $2\times10^{-3}$~a.u. We note in passing that we trained the networks both for spin-restricted and spin-unrestricted DFT with similar results. As such, we decided to use the spin-restricted formalism in the following.

While this strategy resolved, to a large extent, the many-electron self-interaction error, a problem still remains in the vicinity of the integer particle numbers. In fact, our network is fully differentiable, and does not (in fact can not) exhibit a true derivative discontinuity (in a mathematical sense) as a function of $N$.

\subsection{Jump in the xc potential}

We can easily relate the discontinuity of the total energy at integer particle numbers with an uniform shift in the potential. Indeed, the exact uniform shift $\Delta^{N}_{\mathrm{xc}}$ of $v_{\mathrm{xc}}$ at integer particle number $N$ obeys the relation~\cite{Hardy2012}:
\begin{equation}\label{jumpXC2}
\left.\frac{\partial E}{\partial N}\right|_{+}-\left.\frac{\partial E}{\partial N}\right|_{-} = \varepsilon^{N}_{\rm s} + \Delta^{N}_{\rm xc}\,,
\end{equation}
where $\varepsilon^{N}_{\rm s}$ is the Kohn-Sham gap, i.e. the difference between the lowest unoccupied (LUMO) and the highest occupied (HOMO) molecular orbital energies. Noticeably, 
$E(N)$ and $E(N\pm1)$, as well as the eigenvalues corresponding to the LUMO and HOMO can be computed while creating the training sets.

Our second strategy consists of computing, in our learning process, both $\rho_{N + \epsilon}$, and the exact shift $v_{\mathrm{xc}}(N_+)- v_{\mathrm{xc}}(N_-)$. The corresponding mean squared error is then used to extend the loss function in Eq.~\eqref{LossFunction1} 
\begin{equation}
\label{extendedLF}
\mathcal{L} \rightarrow \mathcal{L} + \lambda \,\mathrm{MSE}(v_{\mathrm{xc}}(N+\epsilon) - v_{\mathrm{xc}}(N))\,,
\end{equation}  
where $\lambda$ is an additional hyperparameter.

We expect that this extended loss function can help the functional to learn the correct shift in the derivative. Unfortunately training our basic SWC network with this loss function failed for any learning rate tested.
This can be understood from the fact that our network is still fully differentiable, and can not be forced to learn a discontinuous function. It is clear that to resolve this issue we have to allow explicitly for a discontinuous behaviour in the neural network topology.

\subsection{Incorporating the discontinuity}

\begin{figure}[t!]
\includegraphics[width=0.9\linewidth]{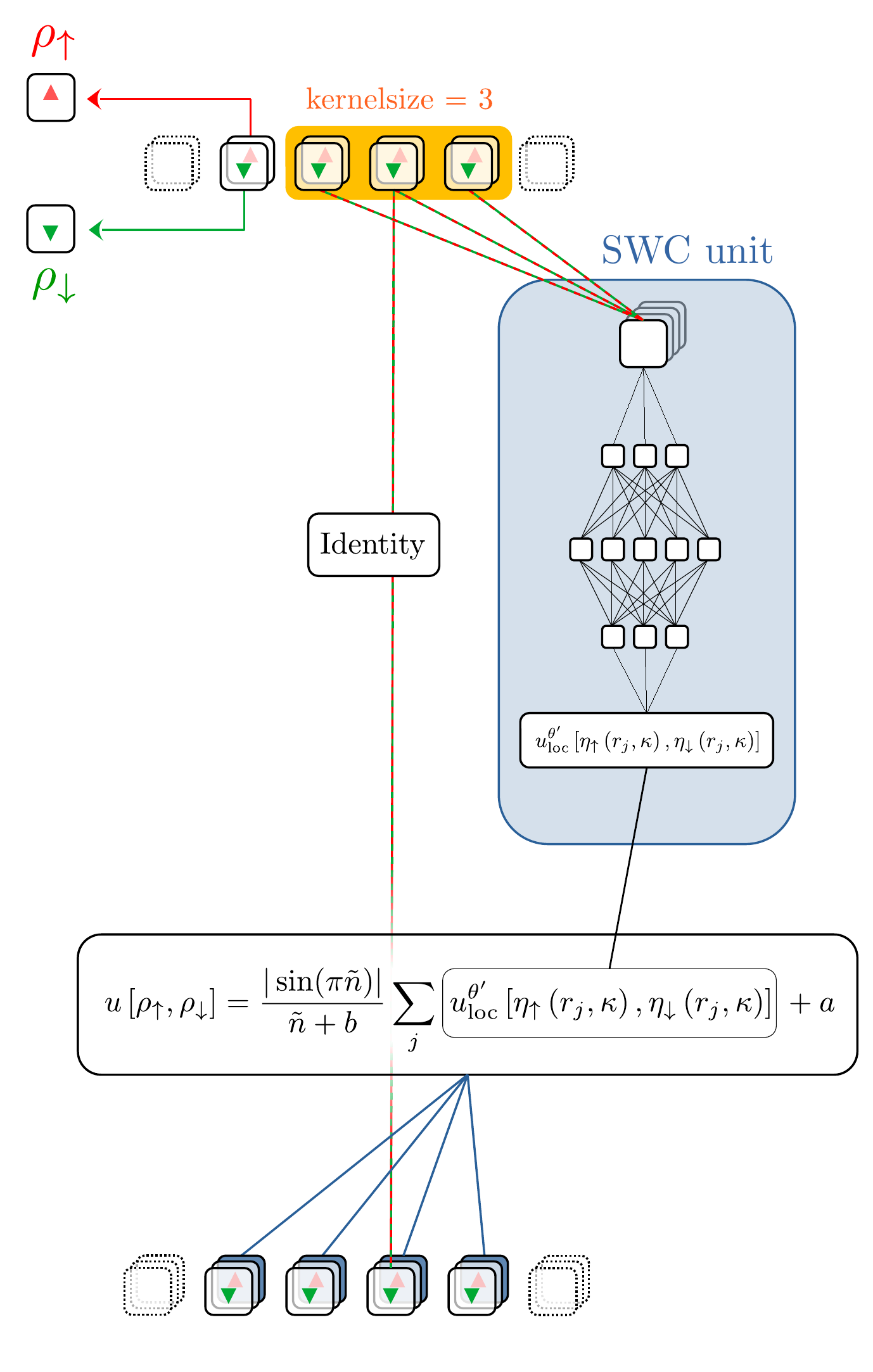}
\caption{\label{AuxFunc}Integration of the non-differentiable auxiliary function $u\left[\rho_{\uparrow}, \rho_{\downarrow}\right]$ as a third channel to the basic SWC unit. The input data contains two channels for both spin-densities $\rho_{\uparrow}$ and $\rho_{\downarrow}$ respectively.}
\end{figure}
An intuitive way to introduce a discontinuity in the derivatives of the
neural network is to use non-differentiable activation functions (e.g., the rectified linear unit~\cite{ReLU}). 
 But there is no obvious reason why a non-differentiability at integer particle numbers will appear -- and these networks will most likely become non-differentiable with respect to the density $\rho$. To overcome this problem we take an alternative route: We define an \textit{auxiliary function} (AF), which we force to be non-differentiable at all integer particle numbers:
\begin{align*}
u\left[\rho_{\uparrow}, \rho_{\downarrow}\right]&= a+
\frac{\left|\sin (\pi \tilde{n})\right|}
{\tilde{n}+b}
\sum_{j}u^{\theta'}_{\mathrm{loc}}
\big[\eta_{\uparrow}(r_j,\kappa), \eta_{\downarrow}(r_j,\kappa)\big] \,,
\end{align*}
where $\tilde{n} = \int (\rho_\uparrow + \rho_\downarrow) \,\mathrm{d}r$ is the (fractional) number of particles and $a$ and $b$ are arbitrary positive constants. Notice that the non-diffe\-ren\-tiability of the AF comes from the non-differentiability of the function $|\sin(\pi \tilde{n})|$.  The local functions 
$u^{\theta'}_{\text {loc }}$ are obtained by using another SWC neural network.
We then replace the functional $E_{\mathrm{xc}}\left[\rho_{\uparrow}, \rho_{\downarrow}\right]$ in Eq.~\eqref{ExcOuput} by $E'_{\mathrm{xc}}[\rho_{\uparrow}, \rho_{\downarrow}] \equiv E_{\mathrm{xc}}\left[\rho_{\uparrow}, \rho_{\downarrow}, u\left[\rho_{\uparrow}, \rho_{\downarrow}\right]\right]$, where the additional channel has been appended to the spin channels.  Here each point carries the (same) information of the value of the non-differentiable AF as illustrated in Fig.~\ref{AuxFunc}. As we show below, this procedure ensures that the machine can learn the non-differentiable function in an efficient manner.

\begin{figure}[t!]
\includegraphics[width=1\linewidth]{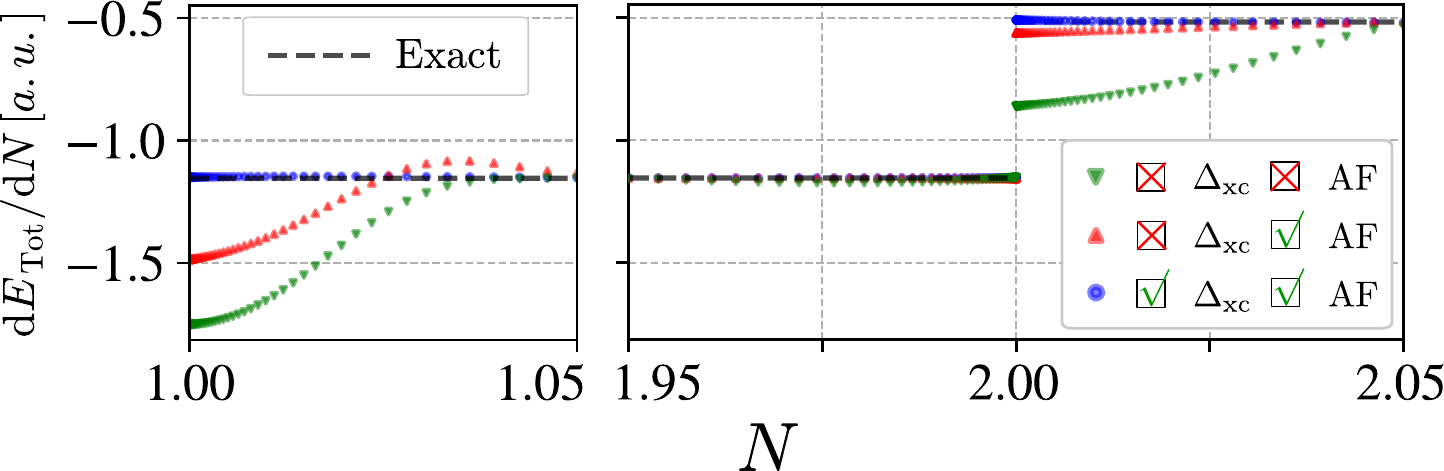}
\caption{\label{Pic:CompareDerivative} Comparison of the derivatives of the energy with respect to the particle number $N$ for three different models evaluated at a randomly chosen external potential: the first model uses only fractional densities, the second incorporates the exact xc shift in the loss function, and the third one uses a non-differentiable AF, in addition to the exact xc shift and the fractional densities.}
\end{figure}

\begin{figure}[t!]
\includegraphics[width=1\linewidth]{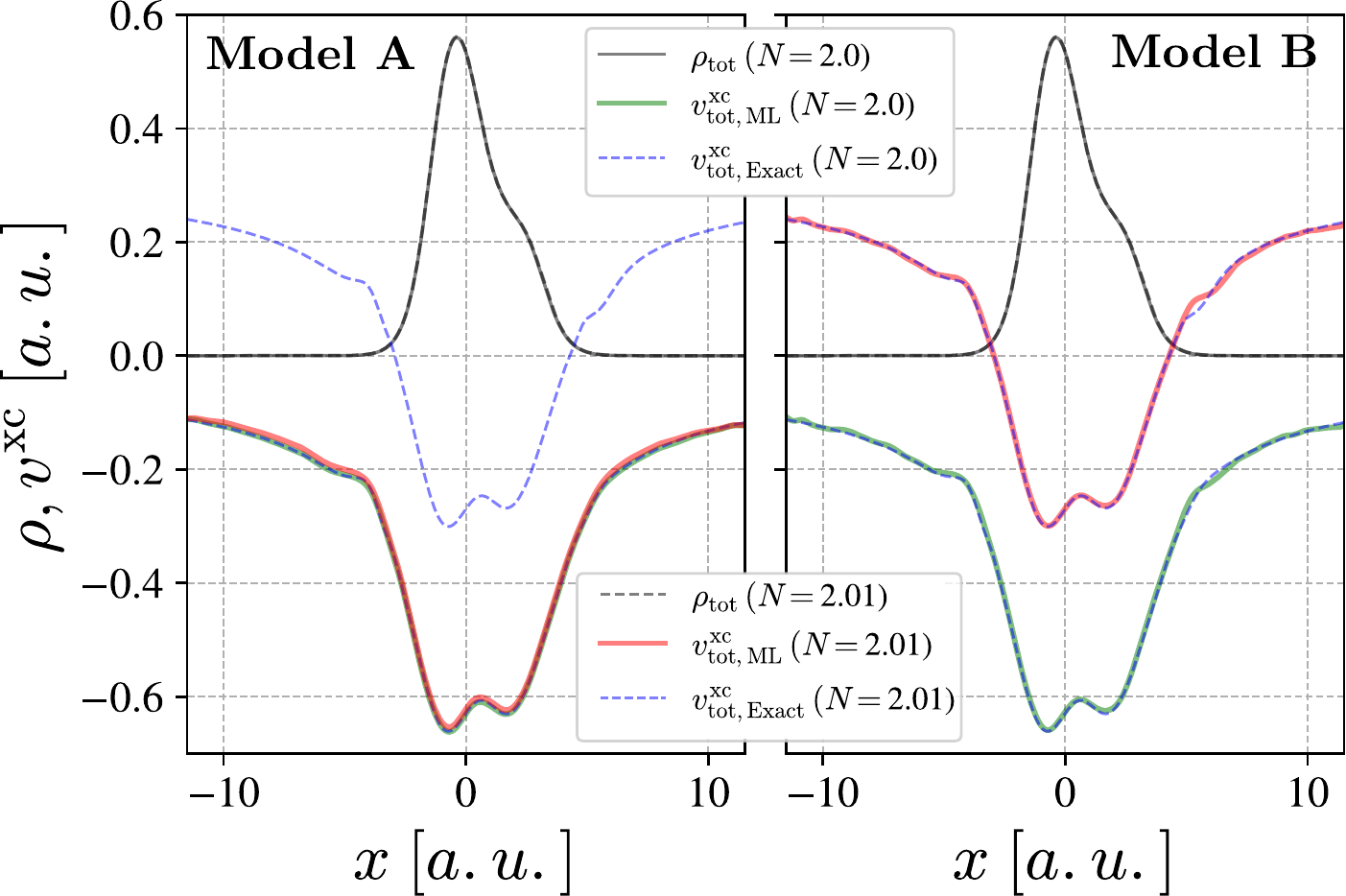}
\caption{\label{Pic:VxcJumpCompare} Comparison between the predicted jump of the xc potential at $N=2$ for Model A and Model B (as explained in the main text). The results correspond to the same external potential used in Fig.~\ref{BestModel}.}
\end{figure}

To study the performance of our approach, Fig.~\ref{Pic:CompareDerivative} presents the predicted derivative of the total energy for three different training models: (1) a model trained only with fractional densities, (2) a model using the AF in the network architecture and trained with fractional densities, and (3) a model using the AF, the shift in the loss function and fractional densities.  First, we should note that the inclusion of the AF solved the training problems discussed in the previous section. As expected, the first model does not predict the correct derivative discontinuity, despite yielding very good errors at fractional particle numbers. While the second model already demonstrates major improvements, especially at $N=2$, our most important finding is that the third model does exhibit a remarkable agreement with the exact results.

In Fig.~\ref{Pic:EnergyCurve} the energy as a function of the particle number is displayed for two models: (A) a model trained with integer densities only and (B) a model using fractional densities, the shift in the loss function and the AF.  The situation is clearly much better for model B, as the energy is very close to linear between integers and shows a cusp at $N=2$. As shown in  Fig.~\ref{Pic:VxcJumpCompare}, this model is also able to reproduce the correct jump in the xc potential: Here we display the xc potentials and densities for a randomly selected external potential at 2  and 2.01 electrons. On the left panel of Fig.~\ref{Pic:VxcJumpCompare} we can see that the xc potential of the basic model A is correct at 2 electrons but barely changes when going from 2 to 2.01. Model B, on the other hand, shows the correct uniform shift in its xc potential.

\section{Conclusions}

We trained a neural network as an exchange correlation functional that (i)~depends explicitly on the number of particles; (ii)~yields total energies that are piece-wise linear between the integers and (iii)~reproduces the infamous derivative discontinuity of the exchange-corre\-lation energy with remarkable  accuracy. To do so we extended the sliding window convolution algorithm to systems with fractional number of particles, and developed a non-differentiable auxiliary function that allows the network to learn correctly the derivative discontinuity.  The most efficient way to train a model yielding highly accurate predictions for the energy in the entire range of particle numbers is by
(i) adding multiple fractional densities into the training data, (ii) training for the correct shift in the xc potential at integer particle numbers, and (iii) incorporating an auxiliary function with a discontinuous derivative with respect to the particle number at integers. 

Our work pushes forward the on-going research on machine-learning functionals by incorporating the correct physics into the training process, thereby improving the path towards an exact functional \cite{Medvedev49}.  As an outlook, we think it will be important to incorporate and test our results for realistic 3D systems. We also expect that our results can stimulate research on similar problems for ensemble DFT (understood as mixtures of ground and excites states), orbital free DFT or functional theories of reduced density matrices \cite{Senjean2018,doi:10.1063/5.0007388,BosonicRDMFT,PhysRevResearch.2.013159,doi:10.1063/5.0023955}.

\section{Methods}

In this section we present all methods and computational details necessary to arrive at our results. First we discuss the generation of the training data and second the details of the network implementation.

\subsection{Exact calculations}

To train and test our models we created a set of 1-di\-men\-sional exact calculations and Kohn-Sham inversions. We sampled 1500 external (Coulomb-) potentials and computed the exact electronic ground state densities as well as the corresponding energies by solving the eigenvalue problem for the electronic Hamiltonian
\begin{multline}
H =-
 \sum_{i=1}^{N} \frac{\mathbf{\nabla}_{i}^{2}}{2}  - \sum_{j=1}^{K} \sum_{i=1}^{N} \frac{Z_{j}}{\sqrt{1 + \left|R_{j}-r_{i}\right|^2}} \\ 
 +\sum_{i<j}\frac{1}{\sqrt{1 + \left|r_{i}-r_{j}\right|^2}}\,,
 \label{EigenvalueProblemShort}
\end{multline}
where $K$ is the total number of nuclei, the variables $R_{j}$ and $Z_j$ denote the position and charge of the $j$th nuclei respectively, $N \in\{1,2,3\}$ is the total number of electrons, and $r_{i}$ is the position of the $i$th electron. We solve the exact ground-state problem with Octopus \cite{octopus}, using a grid spacing of $0.1$ a.u.~and a box size of 23 a.u.~(leading to a grid with 231 points). To circumvent the integrability problem of the Coulomb interaction in 1D we used a softened interaction. The total number of nuclei $K$ were set to be 1, 2 or 3, such that their individual charges satisfy $\sum_{k} Z_k= 3$. Their positions were randomly distributed with $|R_k| \leq 4$ a.u.

\subsection{Spin densities}
As Octopus \cite{octopus} only provides directly spin-densities and wave-functions for two-particle systems we now discuss the problem of obtaining the spin densities for the three particle systems.

We start by noticing that solving the eigenvalue problem $\hat{H}\Phi = E\Phi $ for the Hamiltonian \eqref{EigenvalueProblemShort} yields all many-particle solutions, including both fermionic and bosonic states. A spin-adapted fermionic solution of the form $\Psi\left(r_1 \sigma_1,  \ldots, r_{N} \sigma_{N}\right)$ can be obtained by projecting any spatial solution $\Phi$ on the Young diagrams belonging to certain spin quantum numbers ($S$, $M$), with $S$ being the total spin  and $M = \sum_i \sigma_i$ \cite{spinbook,doi:10.1063/1.4812566}. For a detailed description we refer to \cite{spinbook,WaltersQM,McWeeny}.
Let us denote a set of $f$ primitive, degenerated, but orthogonal spin functions as $\{X(N, S, M ; i)\}_{i=1\cdots f}$. A permutation $\mathbf{P}$ acting on a spin function can be expressed as a linear combination of all primitive spin functions:
\begin{equation}\label{PermSpinEigenfuncs}
\mathbf{P} X(N, S, M ; i)=\sum_{j=1}^{f} X(N, S, M ; j) U(P)_{j i}^{S}\,.
\end{equation}
The expansion coefficients $U(P)_{j i}^{S}$ can be calculated using the orthogonality of $X(N, S, M ; j)$ \cite{Porter}. By taking into account the antisymmetrization 
$\mathcal{A}=\frac{1}{\sqrt{N !}} \sum_{P}(-1)^{P} \mathbf{P}$
of the product of the spin and spatial parts $\Phi X(N, S, M ; i)$ one obtains a sum of products of spatial and spin functions, as follows:
\begin{align}
\Psi_{i}&= \mathcal{A} \Phi X(N, S, M ; i) \nonumber \\ &= 
\frac{1}{\sqrt{N !}} \sum_{P}(-1)^{p} \mathbf{P}^{r} \Phi \mathbf{P}^{\sigma} X(N, S, M ; i)
\nonumber 
\\ &= \label{AnsatzEigenfunction2Short}
\frac{1}{\sqrt{f}} \sum_{j=1}^{f} X(N, S, M ; j) \Phi_{j i}^{S}\,,
\end{align}
where $\mathbf{P}^{r}$ and $\mathbf{P}^{\sigma}$ denote that the permutations operates on spatial and spin coordinates respectively, and
\begin{equation}
\Phi_{j i}^{S}=\sqrt{\frac{f}{N !}} \sum_{P} U(P)_{j i}^{S}(-1)^{p} \mathbf{P}^r \Phi\left(\mathbf{r}_{1}, \ldots, \mathbf{r}_{N}\right)\,. 
\end{equation}
For the ground state of $N=3$ two linear independent spin-eigenfunctions can be chosen:
\begin{subequations}\label{spineigenfuncs_N=3}
\begin{align}
X\left(3, \nicefrac{1}{2}, \nicefrac{1}{2} ; 1\right) & =\frac{1}{\sqrt{6}}\bigl[2\uparrow \uparrow  \downarrow-(\uparrow  \downarrow  \uparrow +\downarrow \uparrow  \uparrow )\bigr] \,,\\ 
X\left(3, \nicefrac{1}{2}, \nicefrac{1}{2} ; 2\right) & =\frac{1}{\sqrt{2}}\bigl(\uparrow  \downarrow \uparrow -\downarrow \uparrow \uparrow\bigr)\,.
\end{align}
\end{subequations}
The (normalized) spatial parts in Eq.~\eqref{AnsatzEigenfunction2Short} and the corresponding spin densities $\rho_{\uparrow}(x)$ and $\rho_{\downarrow}(x)$ can then be found. For instance,
\begin{align}\label{SpinUpDens}
\rho_{\uparrow}(x)  &=
\iint \Bigl[5|\tilde \Phi_{1}|^2 + 9|\tilde \Phi_{2}|^2\Bigr] (\mathrm{d}x_2\mathrm{d}x_3 + \mathrm{d}x_1\mathrm{d}x_3) \nonumber \\ &\qquad \quad +\iint \Bigl[2|\tilde \Phi_{1}|^2 + 18|\tilde \Phi_{2}|^2\Bigr] \mathrm{d}x_1\mathrm{d}x_2  \,,
\end{align}
with $\tilde \Phi_{1} = 2\,\Phi^{S=\nicefrac{1}{2}}_{11}$ and $\tilde \Phi_{2} = 2 \Phi^{S=\nicefrac{1}{2}}_{21}/\sqrt{3}$. 

\begin{figure}[t!]
\centering
  \includegraphics[width=0.9\linewidth]{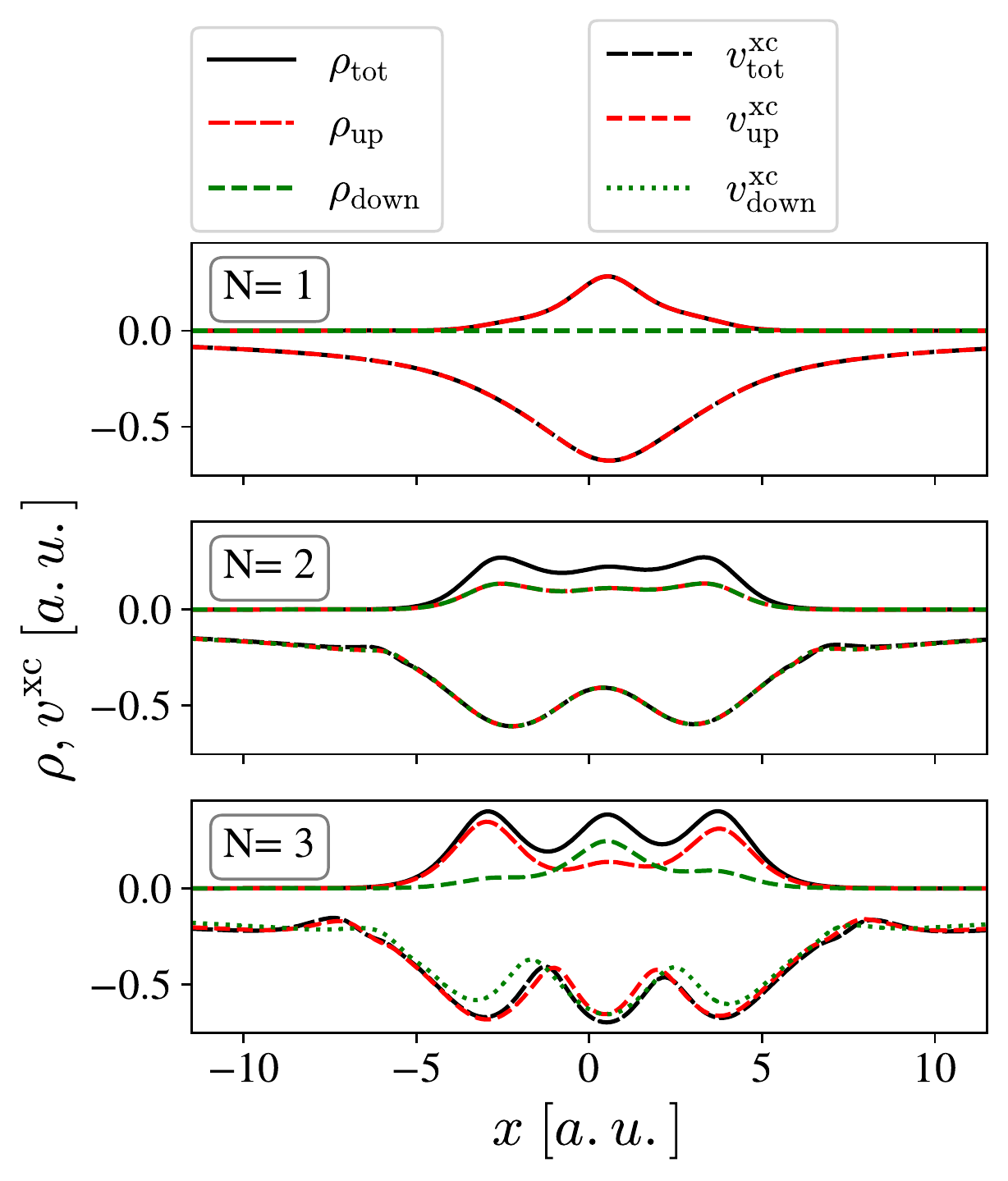}
\caption{Densities and corresponding xc potentials for a certain external potential for different integer particle numbers.}
\label{fig:TriplePlots}
\end{figure}
\begin{figure}[t!]
\centering
  \includegraphics[width=0.9\linewidth]{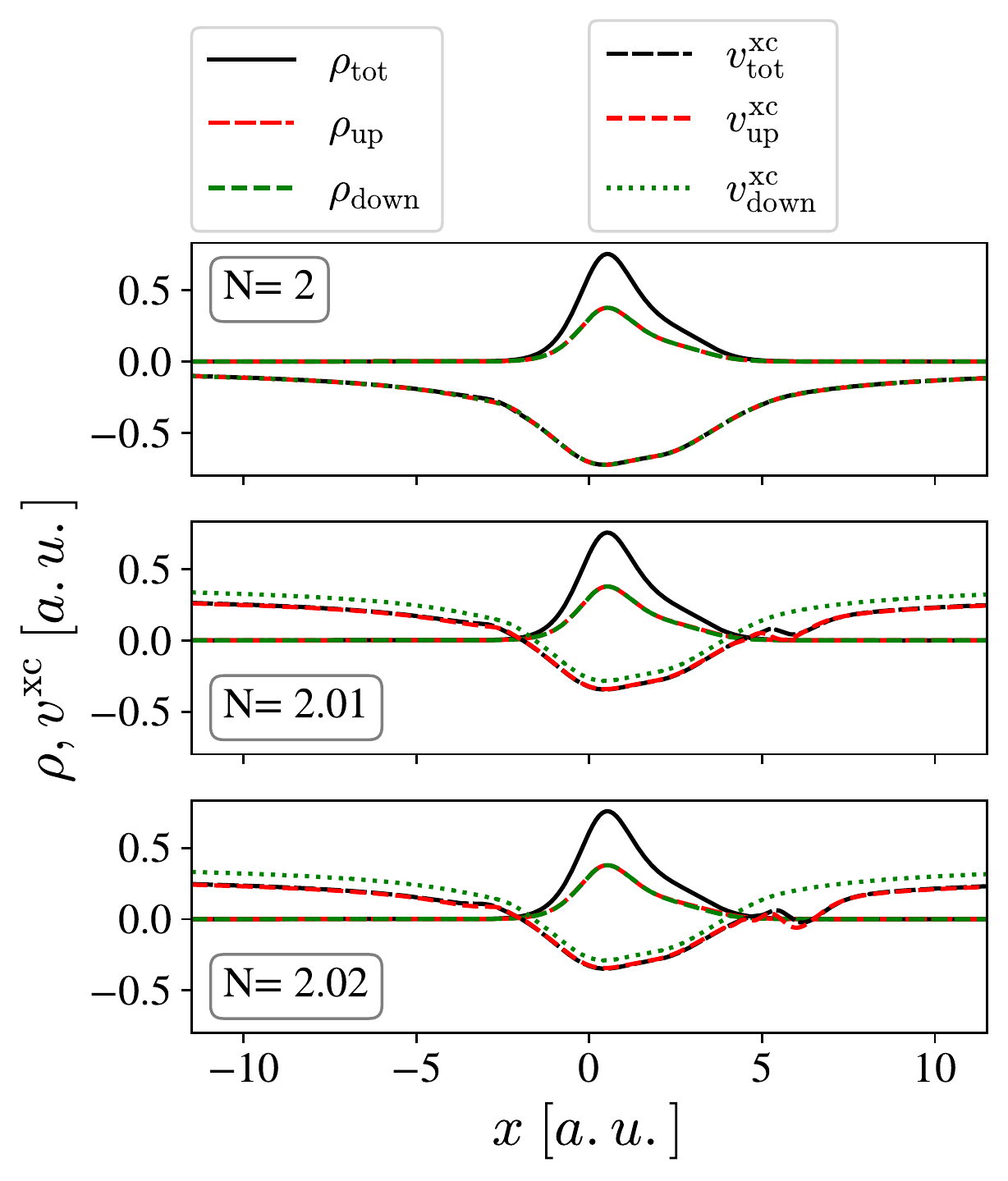}
\caption{Densities and corresponding xc potentials for a certain external potential for different fractional particle numbers.}
\label{fig:TriplePlots2}
\end{figure}

\subsection{Kohn-Sham inversion}

For the inversion of the densities we used the optimization algorithm proposed in Ref.~\cite{Inversion}, that casts the inverse DFT problem of finding the $v_\mathrm{xc}(\rr)$ that yields a given density $\rho(\rr)$ as a constrained optimization problem. The conjugate gradient method was used to update the xc potentials $v_{\mathrm{xc}}^{\sigma}(\rr)$. A constant weight function $w\equiv1$ was also used. 

The densities of  $N=1,2,3$ particles were mixed, allowing us to generate a fractional densities for each external potential. For instance, the set $\{1, 1.5, 2, 2.5, 3\}$ contains additional $\epsilon=0.5$ fractional densities beside the integer ones. If the inversion algorithm did not converge to a MSE below $1.5\times10^{-7}$ for a given density, we removed all samples corresponding to that external potential. The computed xc potentials were shifted by a constant to be in agreement with Koopman's theorem~\cite{KoopmanSpin}.

In Fig.~\ref{fig:TriplePlots} we show a few examples of such spin-den\-si\-ties and inverted xc potentials. For the fractional densities plotted in Fig.~\ref{fig:TriplePlots2} the shift of the xc potential caused by the $\Delta_{\mathrm{xc}}$ jump is clearly visible.

\subsection{Network implementation}

The neural networks were implemented in pytorch \cite{pytorch} using pytorch-lightning \cite{lightning} to simplify the training process.
As discussed before we used the sliding window convolution as proposed in Ref.~\cite{Schmidt2019}.
For each network the number of  zero-paddings  at  the system's boundaries was set  to $(\kappa - 1 ) / 2$ ($\kappa$ is an odd number in our calculations). For all models presented in this paper, we chose a kernel size of $\kappa=201$, corresponding to highly nonlocal funtionals. Each local density was fed into a fully connected network with SILU \cite{Silu} activation functions, and the output layer returned the \textit{local} functions described in the previous sections. The use of SILU activation functions guaranteed the smoothness of the exchange correlation potential and its higher order derivatives. All hidden layer sizes were set to 32. Including the window convolution we used 4 hidden layers.  The additional SWC unit for the AF shared the same hyperparameters. The network weights were optimized with \textsc{Adam} while using a cyclic learning rate scheduler. We used a learning rate of $7\times10^{-4}$ with a batch size of 30. The only exceptions were models trained with the xc-jump in the loss function. In this case each batch contained one  density, corresponding to a random external potential. If the $\Delta_{\mathrm{xc}}$-shift was incorporated in the loss function, the densities per batch consisted of \textit{all} fractional densities of a certain external potential. For these networks we kept the learning rate used before, but changed the batch size to one. The models were trained for $30\,000$ epochs and the model with the best validation loss was selected for testing. 

\bibliographystyle{apsrev4-2}
\bibliography{paperBib}

\section{Competing interests} 

The authors declare no competing interests.

\end{document}